\documentclass[reprint,superscriptaddress,showpacs,amsmath,amssymb,aps,pre]{revtex4-1}
\usepackage{amssymb}
\usepackage{amsmath}
\usepackage{graphicx}
\usepackage{dcolumn}
\usepackage{bm}
\usepackage{hyperref}
\hypersetup{colorlinks=true, citecolor=blue, urlcolor=blue, linkcolor=blue}
\usepackage{amssymb}
\usepackage{color}
\usepackage{graphicx}
\usepackage{amsmath}
\usepackage{mathrsfs}
\usepackage{times}
\usepackage{subeqnarray}
\usepackage{cases}
\usepackage{bm}
\usepackage{diagbox}
\usepackage{booktabs}
\usepackage{bbm}
\usepackage[export]{adjustbox} 
\usepackage[table,xcdraw]{xcolor}
\usepackage{colortbl}
\usepackage{epstopdf}
\definecolor{tabcolor}{rgb}{.105,.410,.113}
\usepackage{array}
\newcolumntype{C}[1]{>{\centering\arraybackslash}p{#1}}

\begin{document}

\title{Evolution of cooperation in a bimodal mixture of conditional cooperators}

\author{Chenyang Zhao}
\affiliation{School of Physics and Information Technology, Shaanxi Normal University, Xi'an 710061, P. R. China}
\author{Xinshi Feng}
\affiliation{McKelvey School of Engineering, Washington University in St. Louis, 63130, United States}
\author{Guozhong Zheng}
\affiliation{School of Physics and Information Technology, Shaanxi Normal University, Xi'an 710061, P. R. China}
\author{Weiran Cai}
\affiliation{School of Computer Science, Soochow University, Suzhou 215006, P. R. China}
\author{Jiqiang Zhang}
\affiliation{School of Physics, Ningxia University, Yinchuan 750021, P. R. China}
\author{Li Chen}
\email[Email address: ]{chenl@snnu.edu.cn}
\affiliation{School of Physics and Information Technology, Shaanxi Normal University, Xi'an 710061, P. R. China}

\begin{abstract}
Extensive behavioral experiments reveal that conditional cooperation is a prevalent phenomenon. Previous game-theoretical studies have predominantly relied on hard-manner models, where cooperation is triggered only upon reaching a specific threshold. However, this approach contrasts with the observed flexibility of human behaviors, where individuals adapt their strategies dynamically based on their surroundings. To capture this adaptability, we introduce a soft form of conditional cooperation by integrating the Q-learning algorithm from reinforcement learning. In this form, players not only reciprocate mutual cooperation but may also defect in highly cooperative environments or cooperate in less cooperative settings to maximize rewards. To explore the effects of hard and soft conditional cooperators, we examine their interactions in two scenarios: structural mixture (SM) and probabilistic mixture (PM), where the two behavioral modes are fixed and probabilistically adopted, respectively.
In SM, hard conditional cooperators enhance cooperation when the threshold is low but hinder it otherwise. Surprisingly, in PM, the cooperation prevalence exhibits two first-order phase transitions as the probability is varied, leading to high, low, and vanishing levels of cooperation. Analysis of Q-tables offers insights into the ``psychological shifts" of soft conditional cooperators and the overall evolutionary dynamics. Model extensions confirm the robustness of our findings. These results highlight the novel complexities arising from the diversity of conditional cooperators.
\end{abstract}
\date{\today }
\maketitle

\section{Introduction}\label{sec1}
Cooperative behaviors are ubiquitous in nature and human societies, underpinning nearly all human activities~\cite{Nielsen1985cooperation,Colman1995Theory, Nowak2006Evolution}. 
Classic economic theory, however, assumes that individuals are rational and self-interested, they make their moves aiming to maximize their rewards. With this logic, cooperators place themselves at a disadvantage, while defection emerges as the more rational choice. This paradox makes the emergence of cooperation difficult to explain and remains a grand challenge till now~\cite {Pennisi2005How}.

To address this puzzle, evolutionary game theory has been employed to study prototypical models such as the Prisoner's Dilemma (PD). In the PD game, two players simultaneously choose whether to cooperate or defect. The core dilemma lies in the fact that, although mutual cooperation yields the highest collective payoff, defection always provides a greater individual payoff, regardless of the opponent's choice. Consequently, a purely rational, payoff-driven player would consistently choose to defect. Numerous mechanisms have been proposed to explain the emergence of cooperation, including kin selection~\cite{Dawkins2006selfish, Wilson1975Sociobiology}, group selection~\cite{Smith1964Group, Charlesworth2000Levels}, direct reciprocity~\cite{Nowak2008Repeated}, indirect reciprocity~\cite{Perc2013Interdependent, Nowak1998indirect, Ohtsuki2006indirect}, network reciprocity~\cite{Nowak1992spatial, Szabo1998Evolutionary, Wang2013Interdependent, Szolnoki2010Dynamically}, dynamical reciprocity~\cite{liang2022dynamical}, and reputation~\cite{xia2023reputation}, among others. Nonetheless, many of these mechanisms remain to be confirmed in experiments~\cite{sanchez2018physics,grujic2014comparative}.

\begin{figure}[t]
\centering
\includegraphics[width=0.95\linewidth]{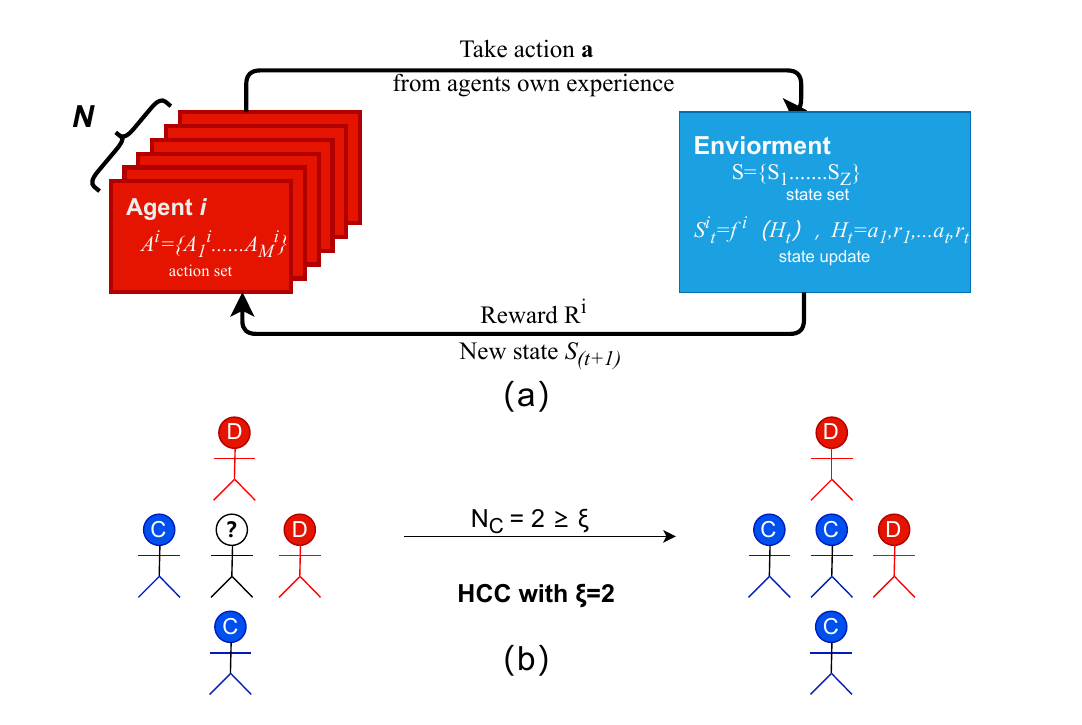}
\caption{\textbf{Two conditional cooperation modes.}
(a) Reinforcement learning (RL) framework acts as the primary behavioral mode, where agents choose actions based on their states, aiming to maximize the accumulated rewards. RL players often behave in the manner of soft conditional cooperation (SCC). (b) Hard conditional cooperation (HCC) mode, where the focal individual counters the number of cooperators in its neighborhood $N_C$ and compares to a threshold $\xi$. Cooperation is chosen only when $N_C\ge \xi$. 
}\label{fig:Scheme}
\end{figure}

In contrast, conditional cooperation (CC)~\cite{Szekely2021Evidence, Devesh2010Conditional, Fischbacher2001Are, Claudia2000The, MartinConditional2008, Thoni2018Ecnomics, MaxwellConditional2016, Wu2014The} has been widely observed in behavioral experiments. For instance, Fischbacher et al.~\cite{Fischbacher2001Are} demonstrated in a public goods game (PGG) that one-third of players are free riders, while approximately half are conditional cooperators. Individuals are classified as conditional cooperators if their contributions increase monotonically with the average contributions of others. Additionally, a detectable proportion of hump-shaped contributors exists, who reduce their contributions when the average contribution of others exceeds a certain threshold. Theoretical studies have shown that conditional cooperators play a crucial role in maintaining and stabilizing cooperation in social dilemmas such as the repeated PGG~\cite{Allen2018Social, Chun2018Endogenous}. This is largely because conditional cooperators can promote sustained cooperation by avoiding exploitation compared to unconditional cooperators. Furthermore, Zhang et al.~\cite{Zhang2021Conditional} found that conditional cooperators do not contribute at higher levels than in a standard PGG but enhance the effectiveness of institutional punishment. However, the conditional cooperation in these studies is typically of a ``hard" nature, where cooperation is adopted only when the number of cooperators in the neighborhood reaches a threshold. This fails to capture the behavioral flexibility and psychological complexities inherent in human decision-making.


Recently, reinforcement learning (RL) has emerged as a new paradigm~\cite{Kaelbling1996Reinforcement, Rangelframework2008, LeeNeural2012, wang2024mathematics} for understanding human behavior, offering a flexible approach to modeling conditional cooperation. Unlike hard conditional cooperators (HCC), RL-based players make decisions guided by policies that are continuously revised to maximize accumulated rewards. In this way, each player develops a unique policy in an introspective manner. To date, several studies have combined RL with evolutionary game theory to investigate various aspects of human behavior~\cite{TomovMulti2021, ZhangOscillatory2020, Wang2022Levy, shengCatalytic2024, JiaLocal2021, D2024emergence}, including cooperation~\cite{Zhang2020Understanding, Zhao2022Reinforcement, Wang2023Synergistic, Ding2023emergence, ZhaoPunish2024, Ma2023emergence, ZHENG2024Evolution, li2025cooperation, mintz2025evolutionary}, trust~\cite{Zheng2023decoding}, fairness~\cite{zheng2024decoding}, resource allocation~\cite{Zhang2019reinforcement, Zheng2023optimal, HeQ2022}, and other collective behaviors~\cite{Zhang2020Oscillatory, D2024emergence}. Notably, Refs.~\cite{ezaki2016reinforcement, horita2017reinforcement} demonstrate that RL explains conditional cooperation and moody conditional cooperation. More recent studies~\cite{YiReinforcement2022, Zhao2024Emergence, shengCatalytic2024} have revealed that RL players exhibit features of conditional cooperation, cooperating in cooperative environments and defecting in defective ones. Interestingly, they also display hump-shaped conditional cooperation, where cooperators switch to defection in highly cooperative neighborhoods to maximize rewards. Given the ubiquity of conditional cooperation and the potential of RL, it is intriguing to incorporate this new RL perspective to describe conditional cooperative behaviors beyond traditional models. The central question we address is: \emph{How does cooperation evolve when diverse conditional cooperation modes are incorporated?}

In this work, we integrate these two types of conditional cooperation (see Fig.~\ref{fig:Scheme}) and investigate the evolution of cooperation by varying their composition. Specifically, we adopt Q-learning as the primary behavioral mode and examine the impact of hard conditional cooperators. This bimodality is implemented in structural, probabilistic, and adaptive manners. We observe diverse dependencies of cooperation on the composition fraction, which can either promote or deteriorate cooperation and exhibit either continuous or discontinuous phase transitions. By monitoring the evolution of Q-tables, we clarify the underlying mechanisms, providing valuable psychological insights into human behavior in real-world scenarios.


\begin{table}[]
\begin{tabular}{c|cc}
\arrayrulecolor{tabcolor}\toprule [1.4pt]
\hline
\diagbox{State}{Action}& C ($a_1$) &  D ($a_2$) \\
\midrule [0.5pt]
\hline
0 $(s_{0})$ & $Q_{s_{0},a_{1}}$ & $Q_{s_{0},a_{2}}$  \\
1 $(s_{1})$ & $Q_{s_{1},a_{1}}$ & $Q_{s_{1},a_{2}}$ \\
2 $(s_{2})$ & $Q_{s_{2},a_{1}}$ & $Q_{s_{2},a_{2}}$ \\
3 $(s_{3})$ & $Q_{s_{3},a_{1}}$ & $Q_{s_{3},a_{2}}$\\   
4 $(s_{4})$ & $Q_{s_{4},a_{1}}$ & $Q_{s_{4},a_{2}}$  \\  
5 $(s_{5})$ & $Q_{s_{5},a_{1}}$ & $Q_{s_{5},a_{2}}$  \\  
\hline
\bottomrule[1.4pt]
\end{tabular}
\caption{Q-table for each Q-learning player. The state $s_{0,...,5}$ corresponds to the number of cooperators in its neighborhood, i.e. the four nearest neighbors plus the player itself, and there are two actions.
}\label{tab:Qtable}
\end{table}
\section{Model}\label{sec2}

Let's consider a population of $N$ individuals playing prisoners' dilemma (PD) game with the four nearest neighbors on a square lattice. The lattice is of size $N=L\times L$ with periodic boundary conditions. In PD, players either cooperate (C) or defect (D). While mutual cooperation yields a reward $R$ for both players, mutual defection results in a punishment payoff $P$; if one cooperates and the other defects, the defector receives the temptation payoff $T$, and the cooperator receives the sucker’s payoff $S$. The payoff matrix is thus as follows:	
\begin{equation}
	\begin{pmatrix}
			\Pi _{CC}  & \Pi _{CD}\\
			\Pi _{DC} &\Pi _{DD}
	\end{pmatrix}=\begin{pmatrix}
			R & S\\
			T & P
	\end{pmatrix}=\begin{pmatrix}
			1 & -b\\
			1+b & 0
	\end{pmatrix}.
	\label{eq:matrix}
\end{equation}
Here, we adopt a strong version of PD, where the four payoffs satisfy $T\textgreater R \textgreater P \textgreater S$. 
$b\in(0,1)$ controls the strength of the dilemma, the larger $b$, the less likely to cooperate.

\begin{figure*}[htbp]
\centering
\includegraphics[width=0.95\linewidth]{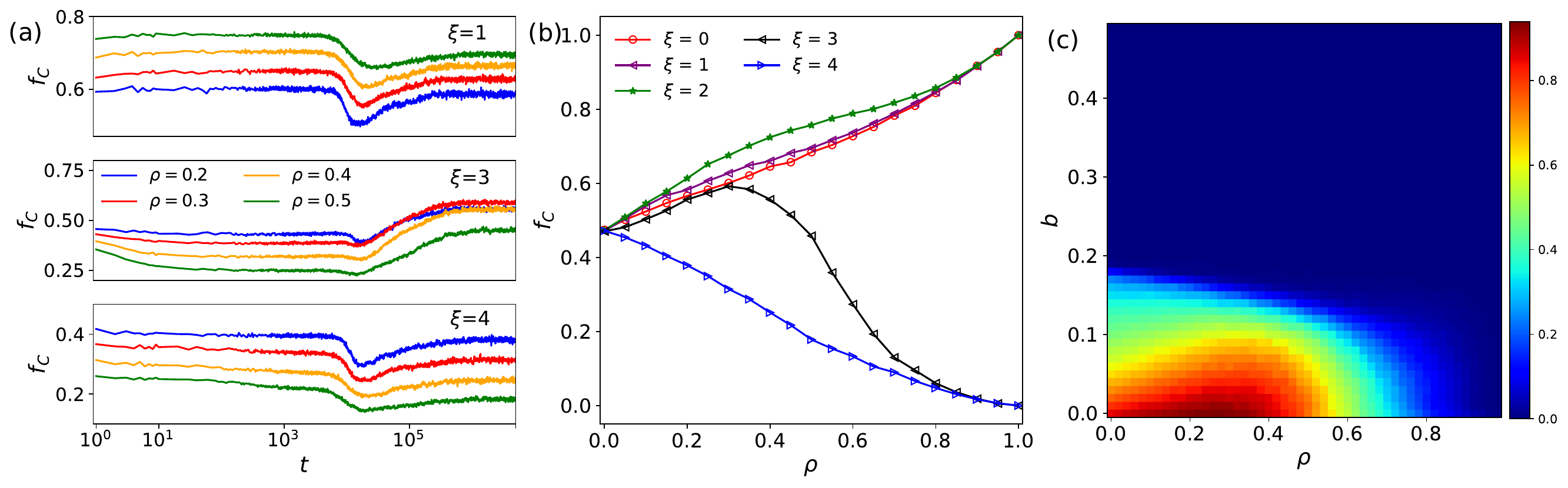}
\caption{\textbf{Evolution of cooperation with structure mixing  (SM).}
(a) Time series of cooperation level $f_C$ for different proportion $\rho$, where the threshold for HCC $\xi=1,3,4$ for the three subplots, respectively. 
(b) The prevalence of cooperation $f_c$ as a function of $\rho$ for $\xi=0,1,...,4$. 
(c) Phase diagram for $f_C$ in the parameter domain of $\rho-b$ with $\xi=3$. 
Other parameter: $b=0.1$ in (a,b).
}\label{fig:Result_SM}
\end{figure*}

In our model, players are assumed to behave within two modes. Players with Q-learning often exhibit a flexible conditional cooperation~\cite{YiReinforcement2022,Zhao2024Emergence,shengCatalytic2024}, acting as the basal mode and is mixed with hard conditional cooperation (HCC), see Fig.~\ref{fig:Scheme}. 
Within the HCC mode,  players cooperate if the number of cooperators in their neighborhood $n_C$ meets the expected threshold $\xi$, i.e. $n_C\ge\xi$, otherwise they defect. Instead, Q-learning individuals make decisions following Q-tables' guidance to maximize their accumulated rewards. Specifically, each player has a Q-table to score values of different actions $a\in \mathbb{A}$ within different states $s\in \mathbb{S}$ (Table ~\ref{tab:Qtable}). The state set $\mathbb{S}=\{s_0,...,s_5\}$ corresponds to the number of cooperators within its neighborhood including itself, and the action set $\mathbb{A}=\{a_1=C, a_2=D\}\equiv\{1, 0\}$. The Q-values $Q_{s,a}$  are the action-values measuring the value of action $a$ within state $s$, the larger the value, the more preferred for the corresponding action $a$ within the state $s$.  

The two behavioral modes are incorporated in two ways: structural mixing (SM) and probabilistic mixing (PM). In SM, each player is assumed to behave in only one mode, where a fraction $\rho$ of people are randomly chosen to behave in HCC mode at the beginning, and the rest adopt Q-learning, and they stick to the chosen modes. Instead, players are allowed to switch their behavioral modes from round to round in PM. Specifically, they behave as a hard conditional cooperator with probability $\rho$ and act according to Q-learning otherwise in each round.

The evolution follows a synchronous updating scheme, and here we first introduce the procedure for the SM scenario. At the very beginning ($t=0$), every player is assigned as a conditional cooperator with probability $\rho$, and behaves according to Q-learning otherwise. Their actions are initialized randomly as either C or D with an equal chance. For those players in the Q-learning mode, the items in Q-table $Q_{s,a}\in(0,1)$ are randomly initialized independently.
At each round $t>0$, if player $i$ is a hard conditional cooperator, she/he makes decisions according to
\begin{equation}
		a_i(t+1) =
		\begin{cases} 
			C,\ n_C(i, t)\ge \xi,  \\
			D,\ n_C(i, t)<\xi, 
		\end{cases}
\end{equation}
\label{eq:hard}
where $n_C(i, t)$ is the number of cooperators in player $i$' neighborhood, and $\xi$ is the expected threshold for cooperation. 

If the player $i$ is within Q-learning mode, she/he makes the decision following the guidance of its Q-table as 
\begin{equation}
	a_i(t+1)=\begin{cases} 
			\arg\underset{a\in A}{\max} Q(s_i(t),a), \ 1-\frac{\epsilon }{2}, \\
			1 - \arg\underset{a\in A}{\max} Q(s_i(t),a),\ \frac{\epsilon }{2}.
		\end{cases}
	\label{eq:maxQ}
\end{equation}
This means that player $i$ chooses the suggested action $\arg\underset{a\in A}{\max} Q(s_i(t),a)$ with a probability of $1-\epsilon$, and 
makes a random exploration with $\epsilon$, where $0<\epsilon\ll1$.

\begin{figure*}[htbp]
\centering
\includegraphics[width=0.95\linewidth]{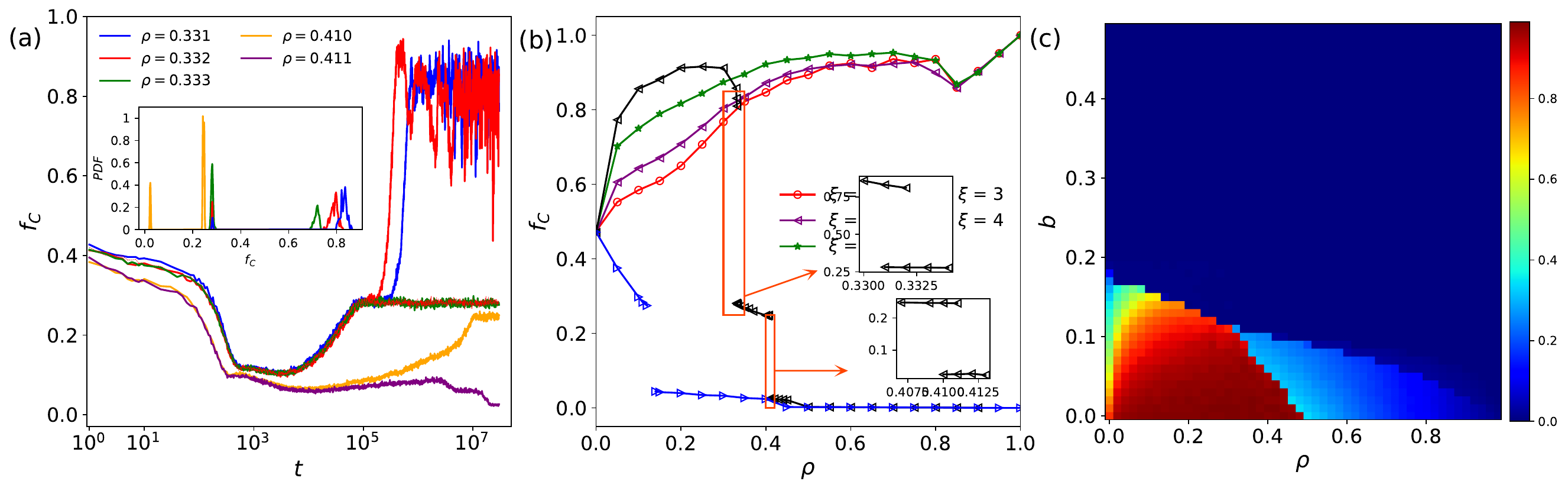}
\caption{\textbf{Evolution of cooperation with probabilistic mixing (PM).}
(a) Time series of cooperation level $f_C$ for different $\rho$, where the threshold $\xi=3$. The insert shows the probability density function (PDF) distribution of $f_C$ within the steady states.
(b) $f_C$ as a function of the probability $\rho$ for $\xi=0,1,...,4$. The two insets show the zoom-in of the two transition regions.
(c) Phase diagram for $f_C$ in the parameter domain $\rho-b$ with $\xi=3$.
Other parameter: $b=0.1$ in (a,b).
}\label{fig:Result_PM}
\end{figure*}

After all the players have made their decisions, their average payoffs can be computed according to the payoff matrix Eq.~(\ref{eq:matrix}), and we have:
\begin{equation}
		\begin{aligned}
			\Pi _{i} =\frac{1}{4} \sum_{j\in e^i}^{}  [Ra_ia_j+Sa_i(1-a_j)+\\ T(1-a_i)a_j+P(1-a_i)(1-a_j)].
		\end{aligned}
	\label{reward}
\end{equation}
The new state is then updated as
\begin{equation}
	s_i(t+1)=a_i(t+1)+\sum_{j\in \Omega_i}^{}\  a_j(t+1),
	\label{State_update}
\end{equation}
where $\Omega_i$ is the neighborhood of player $i$, including its four nearest neighbors. 

After the gaming process is completed, players in Q-learning mode revise their Q-tables to draw lessons at round $t$. Specifically, the Q-values for each player $i$ are updated following the Bellman equation:
\begin{equation}
		\begin{aligned}
		\label{eq:Q_updating}
		Q_{s(t),a(t)}(t\!+\!1)\!=\!(1\!-\!\alpha)Q_{s,a}(t)
		\!+\!\alpha [\Pi_i(t)\!+\!\gamma \max_{a'}Q_{s(t+1),a'}(t)],
		\end{aligned}
\end{equation}
where only the item $Q_{s(t),a(t)}$ in the Q-table is revised, where $\max_{a^{'}}Q_{s(t+1),a'}$ is the maximum reward that can be expected within the new state $s(t+1)$. $\alpha \in(0,1]$ is the learning rate, which captures how fast the old Q-values are revised. Individuals with a small $\alpha$ appreciate their historical experience, and people with a large value are taken as being forgetful. $\gamma \in [0, 1)$ is the discount factor, determining the weight of future rewards. A small value of $\gamma$ means individuals are myopic because they are concerned too much about immediate rewards; instead, players with a larger $\gamma$ are considered farsighted. This then completes the learning process and an elementary step of evolution at round $t$. Repeat the evolution for many rounds when the system reaches a statistically stable state or a predefined time limit.
	 
The evolution procedure remains largely unchanged for the PM scenario, except that the behavioral mode is not fixed but varies. At the beginning of each round $t$, each individual randomly becomes a conditional cooperator with the probability $\rho$; otherwise, she/he falls into the Q-learning mode. In the PM scenario, each player is associated with a Q-table as she/he is now a bimodal player. Notice that the Q-table is revised at each round, even if the player is acting in the HCC mode to mimic the observations of the continuing learning habits of people in their daily lives.

To measure the cooperation level, we define the average cooperation level $f_c$ after transient:
\begin{equation}
		\bar{f_{c}} = \frac{1}{t-t_0} \sum_{\tau =t_0}^{t} f_c(\tau).
	\label{fc}
\end{equation}
Here, $t_0$ is the starting step when the evolution becomes stable.
If not stated otherwise, we fix the game parameter $b=0.1$, the system size $L=100$. For Q-learning mode, the two learning parameters $\alpha = 0.1$, $\gamma = 0.9$, and the exploration rate $\epsilon = 0.01$. For hard conditional cooperators, the expected threshold $\xi=\{0,1,..,4\}$. Each simulation ran for at least $10^6$ transient steps, followed by time averaging over $3\times 10^4$ rounds.	

\section{Result}\label{sec3}
\subsection{Structural mixing}
We first present the results for SM, as shown in Fig.~\ref{fig:Result_SM}, where $\rho$ is the fraction of hard conditional cooperators. 
In one extreme case of $\rho=0$, the population uniformly behaving within Q-learning mode maintains a cooperation level at $f_C\approx0.47$, in line with previous observation~\cite{Zhao2024Emergence}. In the other extreme case $\rho=1$, the resulting cooperation prevalence goes to the two absorbing states $f_C=0$ or 1, depending on the expected threshold $\xi$ and the initial condition. $f_C\rightarrow 1$ for mild expectation $\xi=0,1,2$ and $f_C\rightarrow 0$ for higher values $\xi=3, 4$ for the random initial condition we adopted where $f_C(t=0)\approx 0.5$.

When the hard conditional cooperators are incorporated into the Q-learning population $(0<\rho<1)$, three qualitatively distinct dependencies are observed, the typical time series being as shown in Fig.~\ref{fig:Result_SM}(a). In the cases of $\xi=0,1,2$, the cooperation prevalence $f_C$ monotonically increases with more hard conditional cooperators. The opposite trend is observed in the case of $\xi=4$, where cooperation only occurs within fully cooperative neighborhoods. Interestingly, a non-monotonic dependence is seen for the case of $\xi=3$, where incorporating a few hard conditional cooperators facilitates the cooperation. However, a higher fraction of them damages cooperation, e.g. a smaller $f_C$ is observed for $\rho=0.5$. 
	
The diverse dependencies are more clearly shown in Fig.~\ref{fig:Result_SM}(b), where three different trends of curves are observed.
In particular, the non-monotonic impact of hard conditional cooperators is signified by the peak around $\rho\approx 0.3$. The observation in the case of $\xi=3$ suggests that Q-learning individuals manage to be cooperators by satisfying the requirement of HCC when $\rho$ is small, but their efforts fail when there are too many hard conditional cooperators.  
	
To systematically investigate the impact of HCC as well as the game parameter $b$, a phase diagram is shown in Fig.~\ref{fig:Result_SM}(c), where the cooperation prevalence is color-coded within the parameter domain $\rho-b$ for $\xi=3$. We observe that when the strength of dilemma $b$ is not strong, there is a region where the cooperation prevalence $f_C$ is significantly promoted compared to the benchmark case of $\rho=0$. However, the presence of HCC changes nothing when $b\gtrsim0.18$, and the population falls into the absorbing state of full defection.
	
\subsection{Probabilistic mixing}

When HCC is probabilistically incorporated, the diverse dependencies in SM are maintained but with much more complex properties, as shown in Fig.~\ref{fig:Result_PM}.

Firstly, the presence of mild condition for HCC (i.e. $\xi=0, 1, 2$) also promotes cooperation, but the cooperation prevalence $f_C$ is higher in PM for $\rho\lesssim0.85$ compared to Fig.~\ref{fig:Result_SM}(b). Intuitively, when more players are within the Q-learning mode, they learn that the choice of C brings them a higher payoff in the long term. But if the population is too cooperative ($\rho\gtrsim0.85$), people instead find defection is more profitable when acting within Q-learning mode, as the cooperative environment remains largely unchanged. That explains the slight decrease at the right end of these three curves.

Secondly, the opposite trend is also observed in the case of $\xi=4$, where the HCC mode is supposed to reduce cooperation for random initial conditions. But now the decrease becomes so dramatic, where $f_C$ diminishes in the form of discontinuous phase transition at $\rho\approx 0.13$.

Thirdly and most interestingly, the non-monotonic dependency is also displayed in $\xi=3$, but in the form of double discontinuous phase transitions (the black curve in Fig.~\ref{fig:Result_PM}(b)). When HCC are properly incorporated, the mode mixture promotes the cooperation most significantly compared to cases of $\xi=0,1,2$.
However, at $\rho\approx0.33$, there is an abrupt jump in $f_C$ from a high level of cooperation 0.75 to a low value of 0.25. Further increase in $\rho$ leads the second drop to vanishing cooperation at $\rho\approx0.41$. Typical time series shown in Fig.~\ref{fig:Result_PM}(a), where the evolution of cases with close parameters goes to dramatically distinct destinies. Though there are strong fluctuations observed in high cooperation scenarios (e.g. $\rho=0.332$), this is due to the finite-size effect (Appendix \ref{AppendixA}).
The double peaks of $f_C$ in the inset of Fig.~\ref{fig:Result_PM}(a), together with the bi-stable regions within the two insets in Fig.~\ref{fig:Result_PM}(b) confirm that both phase transitions are first-order.

Fig.~\ref{fig:Result_PM}(c) shows the phase diagram for $\xi=3$ in the parameter domain $\rho-b$. As seen, the discontinuous nature of phase transitions is signified by the sharp boundaries compared to Fig.~\ref{fig:Result_SM}(c). Detailed examination shows that when $b\lesssim0.12$, cooperation undergoes two discontinuous phase transitions as a function of $\rho$. 
But when $0.12\lesssim b\lesssim0.18$, there is only one phase transition, but still of first-order.
For $b\textgreater 0.18$, no phase transition is expected as the population falls into all defection. 
Notice that, by fixing $\rho$, similar cooperation phase transitions are also expected as a function of the game parameter $b$.

\section{Mechanism analysis}\label{sec4}
\subsection{Structural mixing}\label{sec4.1}
\begin{figure}[tbp]
\centering
\includegraphics[width=1.0\linewidth, center]{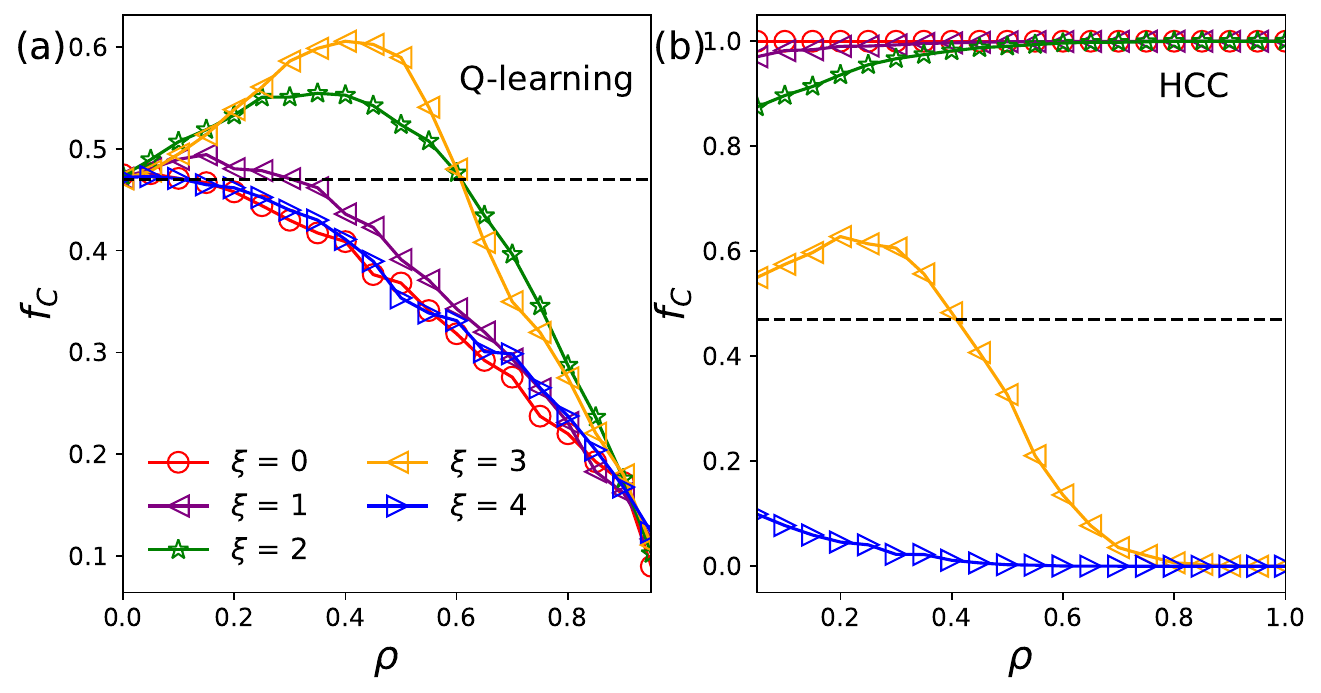}
\caption{\textbf{Separate dependence of cooperation with SM.}
Cooperation level $f_C$ as a function of proportion $\rho$ for different $\xi$ for players within Q-learning (a) and HCC modes (b), respectively. The black dashed line corresponds to the cooperation level for the case with pure Q-learning mode (i.e. $\rho=0$) for comparison. 
}\label{fig:details_SM}
\end{figure}

To understand the mechanism behind the evolution of cooperation within SM, Fig.~\ref{fig:details_SM}(a,b) further provides the detailed dependence of $f_C$ on the proportion $\rho$ for players within the two modes, respectively.
 As shown, players within HCC mode opt for cooperation for mild conditions ($\xi=0,1,2$) across the whole proportion range. As a response, Q-learning players tend to be more defective than the benchmark scenario (i.e. $\rho=0$) for $\xi=0,1$, and the cooperation prevalence $f_C$ continues to decrease as more people are within CC mode. This is not the case for $\xi=2$, where $f_C$ increases first for small $\rho$ and declines after $\rho\gtrsim0.4$. Q-learning players seemingly learn that they have to be more cooperative to trigger players in CC mode to cooperate when $\rho$ is small. Yet, they respond similarly to a large $\rho$ as in cases of $\xi=0,1$ where hard conditional cooperators almost certainly cooperate.

 This non-monotonic dependence is also seen for players in both modes for $\xi=3$. In this case, establishing stable cooperation is challenging from the random initial condition as the condition $n_C\ge 3$ is less likely to be satisfied. Interestingly, Fig.~\ref{fig:details_SM}(a) shows that Q-learning players act in a highly cooperative manner aiming for higher long-term rewards. These efforts successfully maintain a decent level of cooperation for the population with a small $\rho$. However, when hard conditional cooperators dominate (say $\rho>0.4$), the effort of Q-learning players fails to meet the expected threshold. This is because players within CC mode are often surrounded by the same kind of players and they defect with each other. As a result, $f_C$ decreases with increasing proportion $\rho$, and as a response Q-learning players turn to defect more frequently as no higher rewards can be expected if still being cooperators.
 
Finally, for the threshold of $\xi=4$, players within HCC mode opt for defection as the threshold is even harder to meet. As a result, Q-learning players also become more defective as well. With the increasing $\rho$, Q-learning players become increasingly defective, ultimately leading to a vanishing cooperation level.

\begin{figure}[tbp]
\centering
\includegraphics[width=1.0\linewidth, center]{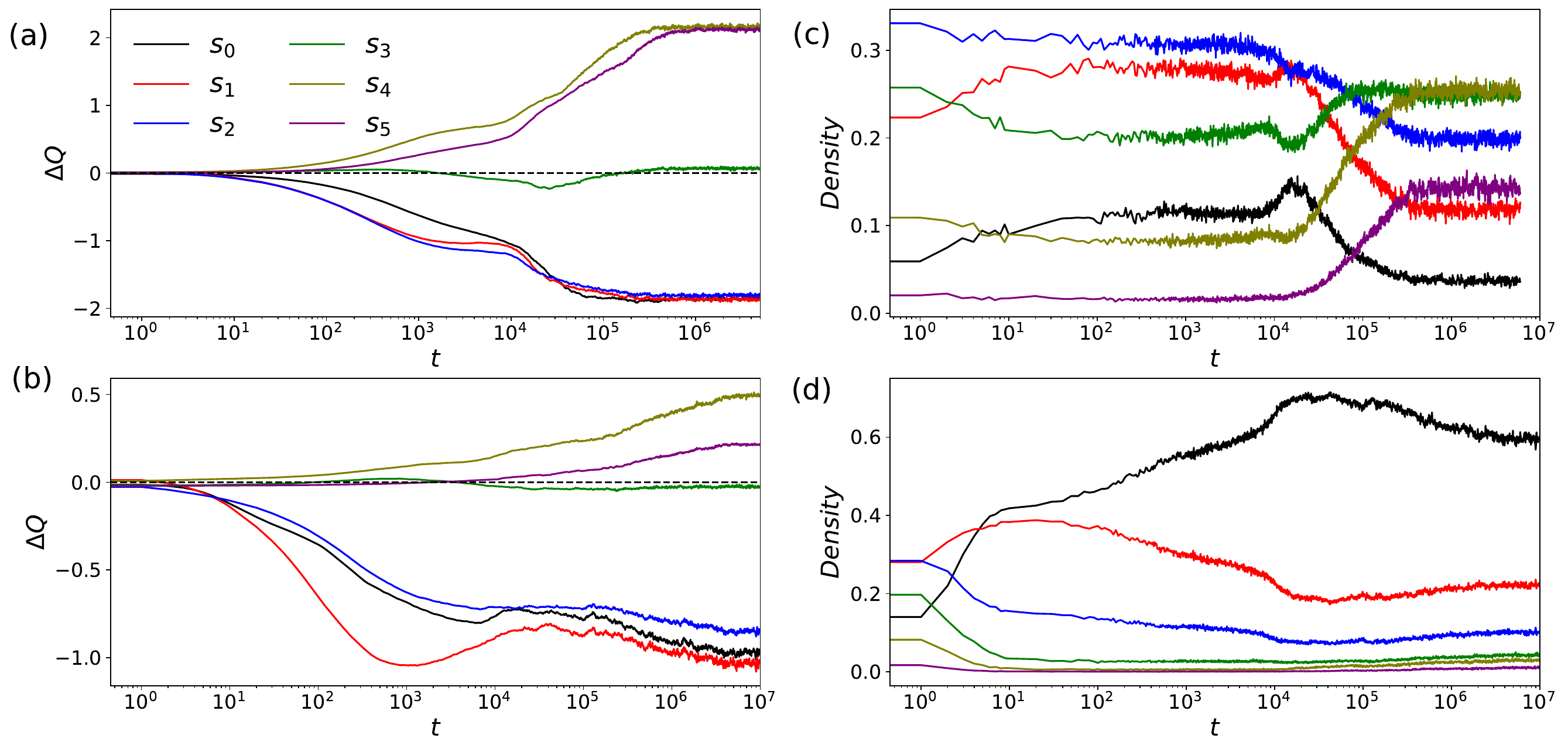}
\caption{\textbf{Evolution of Q-tables and state densities with SM.}
(a, b) The temporal evolution of averaged Q-value difference $\overline{\Delta Q}_{s_{j}}$ defined in Eq.~(\ref{eq:DeltaQ}) for all six states. 
(c, d)  The temporal evolution of all six state densities $f_{s_{0,...,5}}$.
$\rho=0.3$ in (a,c), and $\rho=0.7$ in (b,d),
}\label{fig:Q_table_SM}
\end{figure}
	
\begin{figure*}[htbp]
\centering
\includegraphics[width=1.0in]{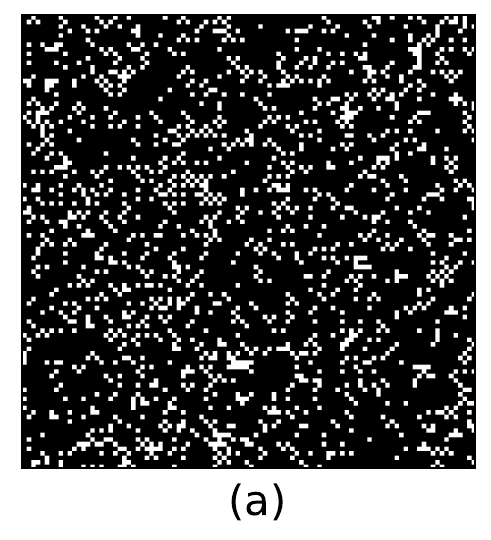}
\includegraphics[width=1.0in]{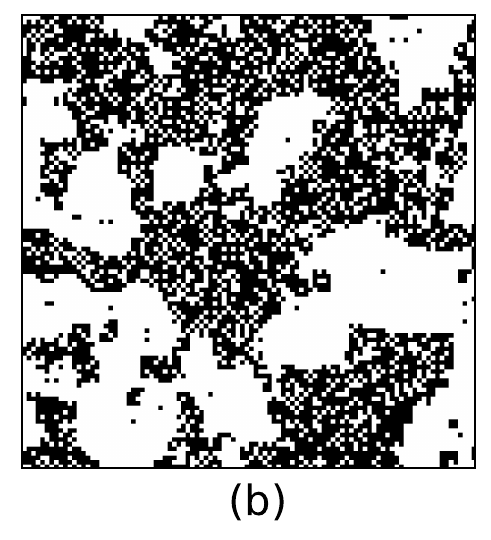}
\includegraphics[width=1.0in]{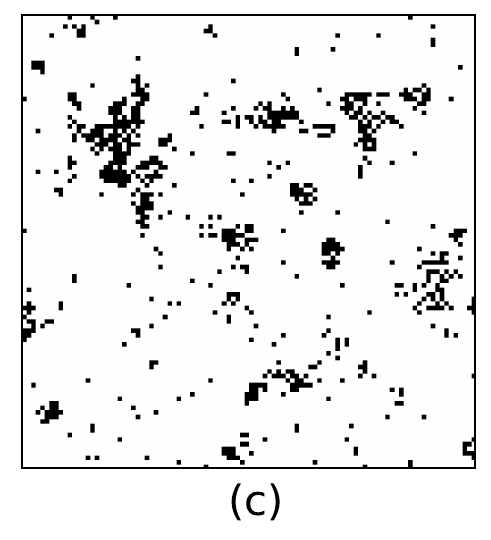}\\
\includegraphics[width=1.0in]{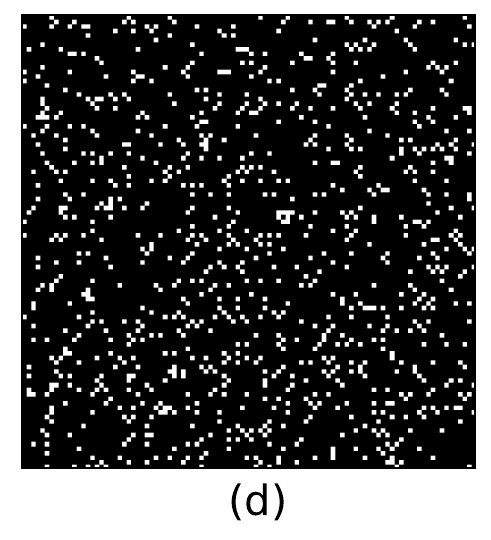}
\includegraphics[width=1.0in]{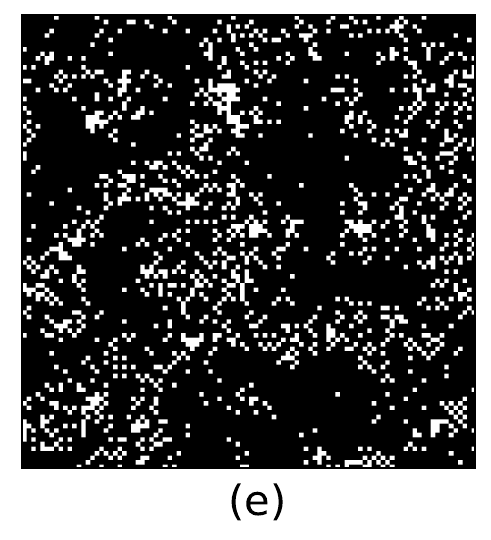}
\includegraphics[width=1.0in]{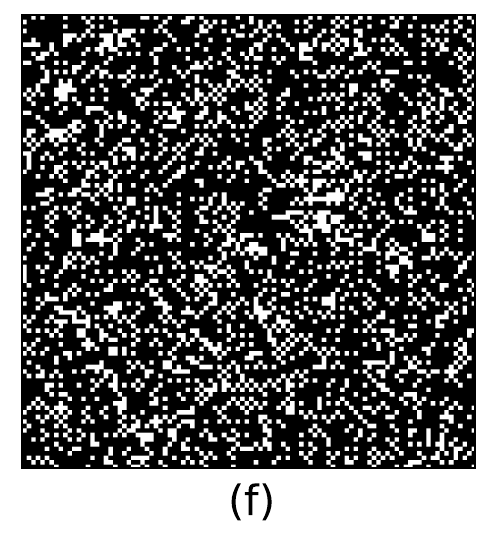}\\
\includegraphics[width=1.0in]{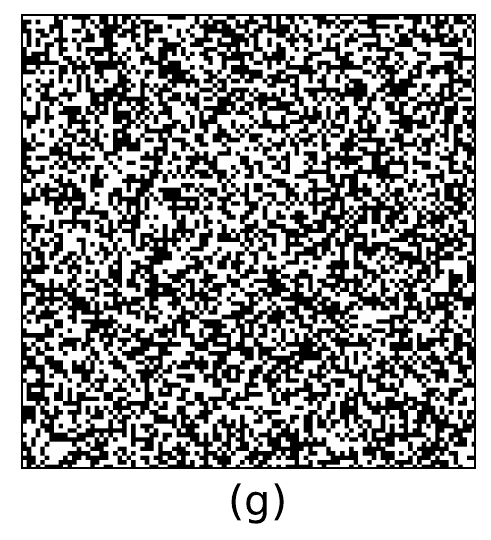}
\includegraphics[width=1.0in]{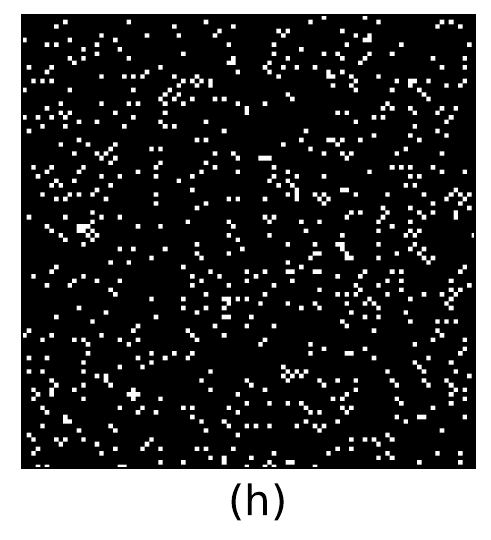}
\includegraphics[width=1.0in]{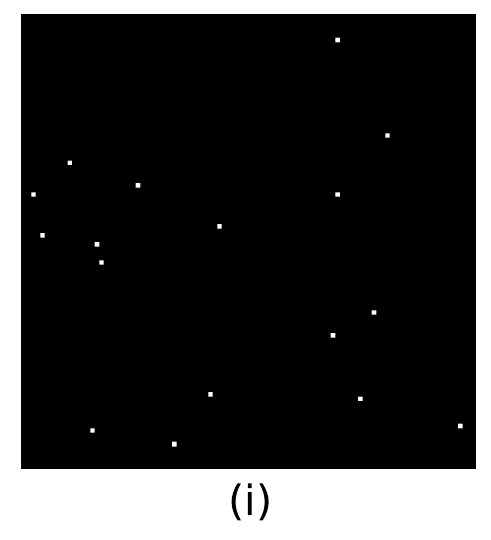}
\caption{\textbf{Typical snapshots with PM.} The cases with $\xi=3$ for $\rho=0.3, 0.4, 0.5$ correspond to the three rows for high, low, and vanishing cooperation levels, respectively. 
Cooperators and defectors are color-coded in white and black. 
(a-c) show evolution to a high level of cooperation with $\rho=0.3$ for $t=10^4$, $8\times10^4$, and $2\times10^5$.
(d-f) are for a low cooperation with $\rho=0.4$ for $t=10^3$, $3\times10^4$, and $1.9\times10^5$.
(h-i) are for the vanishing cooperation with $\rho=0.5$ for $t=0$, $10^3$, and $4\times10^5$. 
The system size is $100\times100$.
}\label{fig:snapshot_PM} 
\end{figure*}

To further understand how Q-learning players make their moves, we define the cooperation preference as follows:
\begin{equation}
	\overline{\Delta Q}_{s_{j}} = \frac{1}{N} {\sum_{i=1}^{N}} (Q_{s_j,C}^{i}-Q_{s_j,D}^{i}),
	\label{eq:DeltaQ}
\end{equation}
where $s_j\in S$ denotes different states and $i$ is the player's labeling. $\overline{\Delta Q}_{s_{j}}>0$ means that the action of cooperation is preferred on average within state $s_j$ in the population, otherwise defection is favored. 

With this in mind, let's monitor the evolution of $\overline{\Delta Q}_{s_{j}}$ as well as the state densities for the case with $\xi=3$, and we focus on two typical scenarios ($\rho=0.3$ and 0.7, see Fig.~\ref{fig:Q_table_SM}), which correspond to high and low cooperation prevalences, respectively.
Interestingly, in both scenarios, the Q-learning players reciprocate cooperative neighborhood with cooperation ($\overline{\Delta Q}_{s_{4,5}}>0$) but defect against defective surroundings ($\overline{\Delta Q}_{s_{0,1,2}}<0$), as shown in Fig.~\ref{fig:Q_table_SM}(a, c). This feature was also reported previously in Refs.~\cite{Zhao2024Emergence,shengCatalytic2024}. A subtle difference in the two cases is that the preference for state $s_3$ is changed from cooperative ($\overline{\Delta Q}_{s_{3}}>0$ for $\rho=0.3$) to defective ($\overline{\Delta Q}_{s_{3}}<0$ for $\rho=0.7$) as the cooperation is worsened.
	
Given these cooperation preferences, the outcome of cooperation is then determined by the states in which players are within. Fig.~\ref{fig:Q_table_SM}(b, d) show that the state density profiles are very different.  For $\rho=0.3$, the density ranking is $d_{s_3}\approx d_{s_4}>d_{s_2}>d_{s_5}>d_{s_1}>d_{s_0}$, while the ranking becomes $d_{s_0}> d_{s_1}>d_{s_2}>d_{s_3}>d_{s_4}>d_{s_5}$ in the case of $\rho=0.7$. This explains why decent cooperation is observed in the former as the cooperative surroundings ($s_{3,4,5}$) are dominating, and also the low cooperation in the latter case as the defective states ($s_{0,1,2}$) abound.
	
\subsection{Probabilistic mixing}\label{sec4.2}
To develop some intuition of the discontinuous phase transitions observed in PM, we first present some typical snapshots for $\xi=3$, as shown in Fig.~\ref{fig:snapshot_PM}. 
Fig.~\ref{fig:snapshot_PM}(a-c) illustrate the evolution towards a high cooperation, while Fig.~\ref{fig:snapshot_PM}(d-f) and~\ref{fig:snapshot_PM}(g-i) depict the cases of a low and vanishing cooperation state, respectively. 
These three spatiotemporal patterns present qualitatively distinct properties.

As shown in Fig.~\ref{fig:Result_PM}(a), the typical time series shows that cooperation declines in all cases initially,  and this is because defection is a more profitable choice in randomly configured surroundings. When $\rho$ is small (e.g., $\rho=0.3$), Q-learning players opt for cooperation as guided by long-term rewards to meet the requirements of HCC players. Once this effort succeeds somewhere, the Q-learning players are reinforced to stick to cooperation, and the condition for players within HCC mode is more likely to be satisfied. This then forms a positive feedback that explains the formation of cooperation clusters shown in Fig.~\ref{fig:snapshot_PM}(a-c), and the system evolves into a high cooperation state as the cooperation clusters expand to take over the whole domain.

However, when the probability acting in HCC mode is not that small (e.g., $\rho=0.4$), the effort of Q-learning players fails to trigger the cooperation propensity of HCC players. Instead, Q-learning players turn to defection, strengthening the defection of HCC players, and defective clusters are formed, as seen in Fig.~\ref{fig:snapshot_PM}(e). However, players within Q-learning mode still adopt cooperation from time to time for higher long-term rewards, their efforts are sometimes rewarded by forming small cooperation clusters, wherein the condition of HCC mode is satisfied, as seen in Fig.~\ref{fig:snapshot_PM}(f).

 Nevertheless, as $\rho$ continues to rise, the efforts of Q-learning players ultimately fail to induce the HCC players, and no cooperation cluster is seen in Fig.~\ref{fig:snapshot_PM}(h). As a result, Q-learning players give up the choice of cooperation, and cooperation is vanishing [Fig.~\ref{fig:snapshot_PM}(i)]. 

\begin{figure}[tbp]
\centering
\includegraphics[width=1.0\linewidth, center]{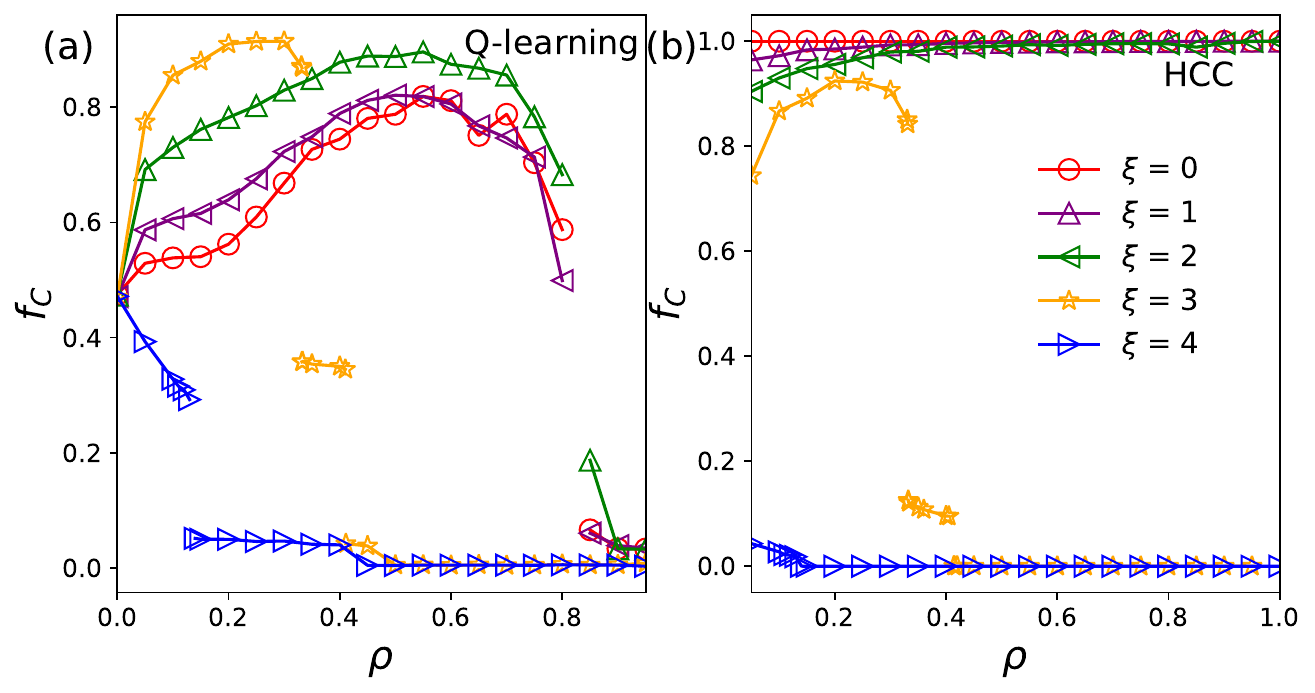}
\caption{\textbf{Seperate dependence of cooperation for the two modes with PM.}
Cooperation level $f_C$ as a function of proportion $\rho$ for different $\xi$ for players within Q-learning (a) and hard conditional cooperation modes (b), respectively. 
}\label{fig:details_PM}
\end{figure}

\begin{figure}[htbp]
\centering
\includegraphics[width=0.95\linewidth]{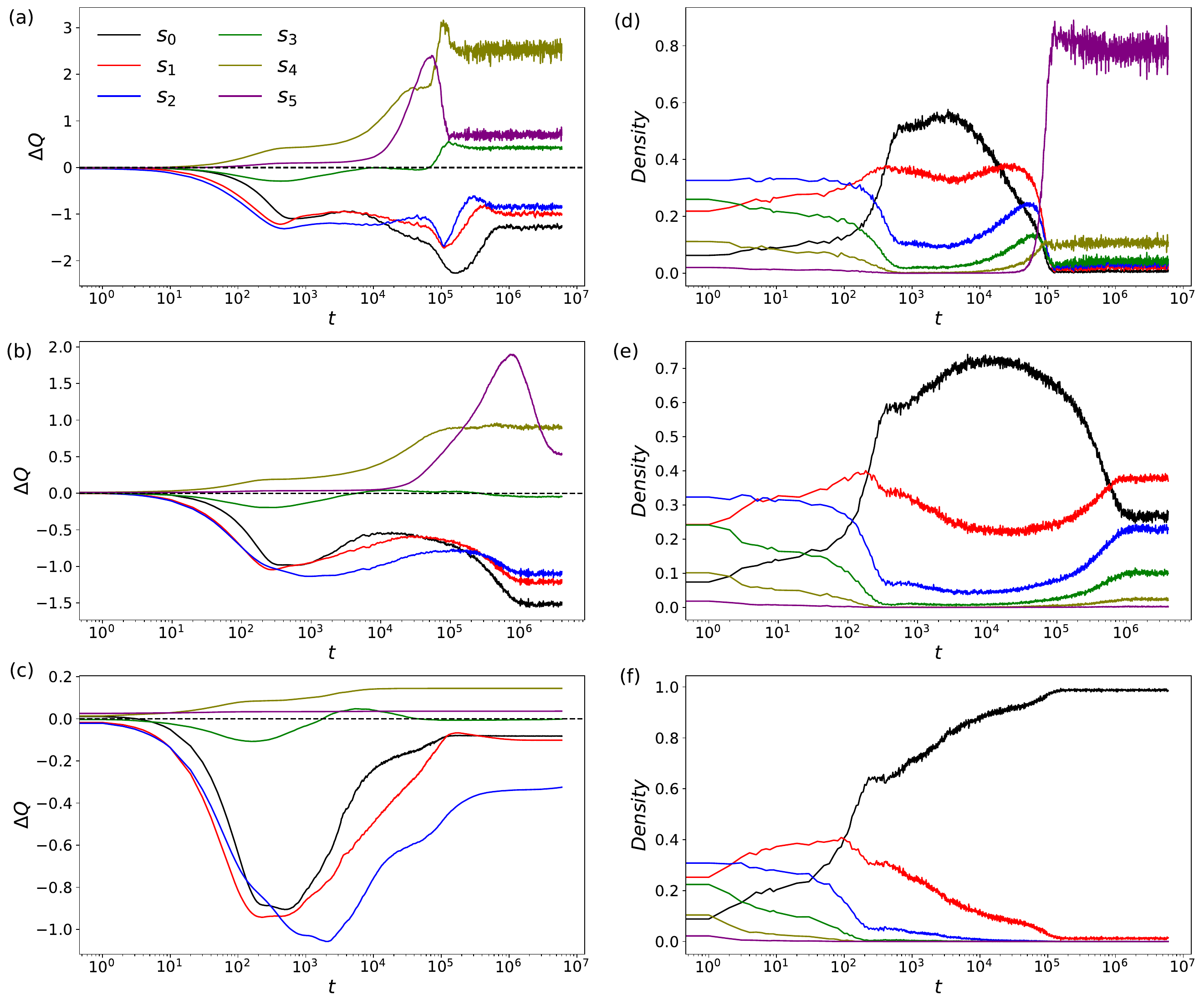}
\caption{\textbf{Evolution of Q-tables and state densities in PM.}
Left column:  The temporal evolution of averaged Q-value difference $\overline{\Delta Q}_{s_{j}}$ defined in Eq.~(\ref{eq:DeltaQ}) for all six states. 
Right column:  The temporal evolution of all six state densities $f_{s_{0,...,5}}$. Parameter for the three rows: (a, d) $\rho=0.3$, (b, e) $\rho=0.4$, (c, f) $\rho=0.5$.
Other parameter: $\xi=3$.
}\label{fig:Q_table_PM}
\end{figure}

\begin{figure}[!tb]
\centering
\includegraphics[width=0.95\linewidth]{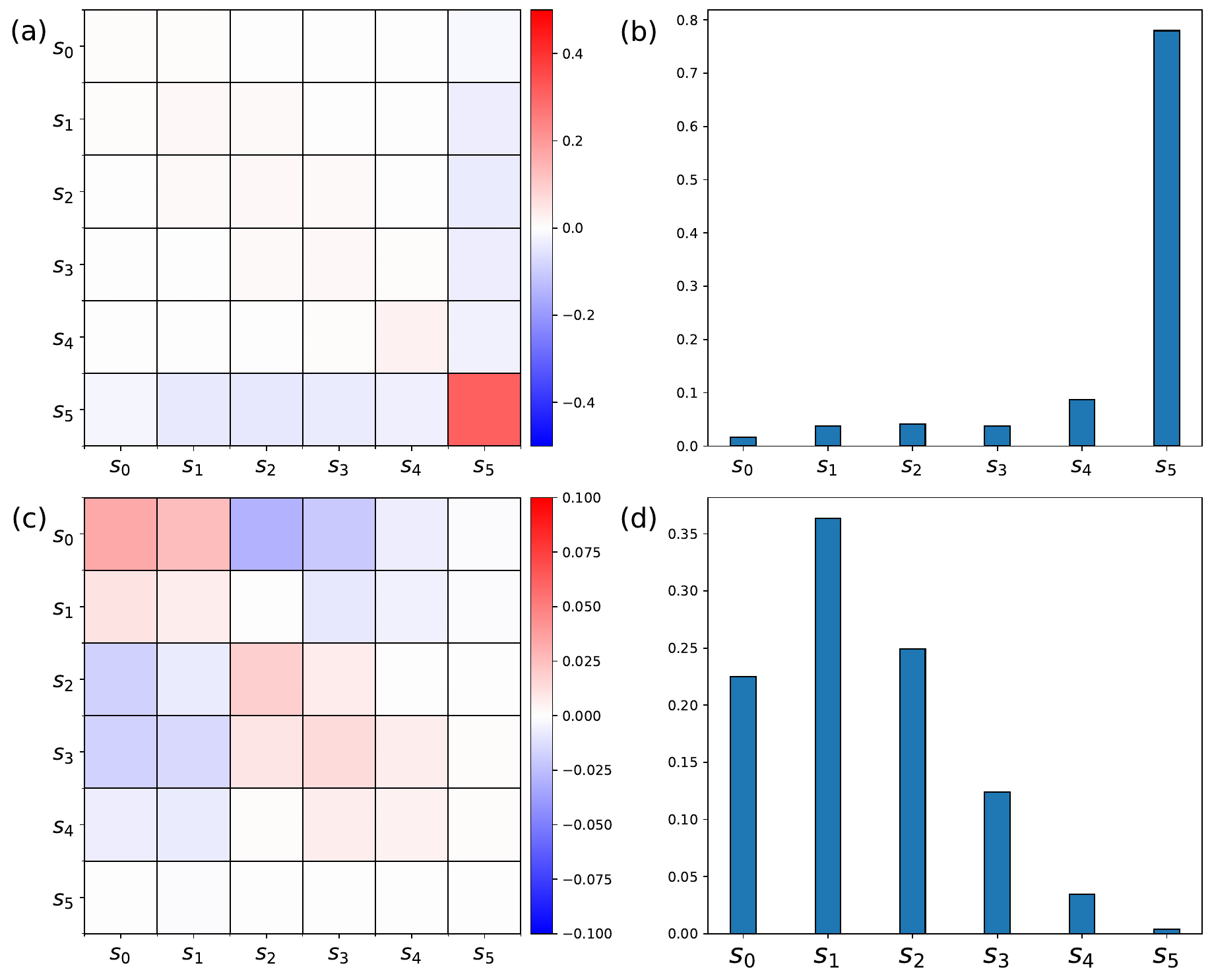}
\caption{\textbf{Matrices of Cohen's kappa coefficients and state distributions.} 
The parameter $\rho= 0.332$ is located within the bistable region, the top and bottom row are for the high and low cooperation, respectively. 
(a,c) show the Cohen's kappa coefficient matrices for the temporal correlation.
(b,d) show the time-average state density $\bar{f}_{s}$ distribution.  
Other parameter: $\xi =3$.
}\label{fig:Kappa_density}
\end{figure}

To better understand the emergence of cooperation with PM, we also provide the separate dependence of cooperation prevalence for the two modes in Fig.~\ref{fig:details_PM}. Compared to the case of SM shown in Fig.~\ref{fig:details_SM}, players in HCC mode also exhibit a high cooperation level for $\xi=0, 1, 2$, and a vanishing prevalence for $\xi=4$. The cooperation prevalence in Q-learning mode presents a non-monotonic trend, but the decay occurs much later and is much sharper compared to Fig.~\ref{fig:details_SM}(a). This also explains the profile valley around $\rho=0.85$ in Fig.~\ref{fig:Result_PM}(b).
More complex dependence happens for $\xi=3$ in the form of double jumps, where the cooperation prevalence in Q-learning mode is higher when $\rho$ becomes large since players in Q-learning mode aim to uphold cooperative surroundings by being cooperators. Their efforts cease when full defection is reached when $\rho\gtrsim 0.4$.

Fig.~\ref{fig:Q_table_PM} shows the evolution of the Q-table with $\xi=3$ for $\rho=0.3, 0.4, 0.5$, corresponding to the three cases shown in Fig.~\ref{fig:snapshot_PM}. The cooperation preference shown in the left column presents that Q-learning players cooperate within a cooperative neighborhood ($\overline{\Delta Q}_{s_{4,5}}>0$), and defect against a defective neighborhood ($\overline{\Delta Q}_{s_{0,1,2}}<0$), in line with the observation in SM. A careful comparison shows that the cooperation preference in state $s_3$ is changed from cooperation ($\overline{\Delta Q}_{s_{3}}>0$ for $\rho=0.3$) to defection ($\overline{\Delta Q}_{s_{3}}<0$ for $\rho=0.4$ and 0.5). 
This suggests that as cooperation decreases, the tolerance of Q-learning agents is also reduced, ultimately leading to a decline and even collapse of the system's cooperation level. Together with Fig.~\ref{fig:Q_table_SM}(a), these observations consistently show that the cooperation preference for Q-learning players is not fixed, they are modified based on their surroundings.

The density evolution shown in Fig.~\ref{fig:Q_table_PM}(d-f) shows their rank changes dramatically for the three cases. The dominating densities are $d_{s_5}$ for $\rho=0.3$, $d_{s_{1,0,2,3}}$ for $\rho=0.4$, and $d_{s_0}$ for $\rho=0.5$, respectively. Combined with the cooperation prevalence shown in Fig.~\ref{fig:Q_table_PM}(a-c), this explains the emergence of high, low, and vanishing cooperation.

To further inspect the state transition, we compute the Cohen's kappa matrix, where the element $\kappa (s,s')$ is the Cohen's kappa coefficient for the temporal correlation between two consecutive states $s$ and $s'$, defined as follows:
\begin{equation}
	\kappa (s,s' ):=\frac{\bar{f}(s,s')- \bar{f}(s)\bar{f}(s')}{1-\bar{f}(s)\bar{f}(s')},
\label{kappa} 
\end{equation}
where $\bar{f}_s$ is the time-averaged state density, and
\begin{equation}
	\bar{f}\left(s, s^{\prime}\right):=\frac{\sum_{\tau=t_{0}}^{t-1} \sum_{i=1}^N \mathbbm{1}_{s^{i}(\tau)=s, s^{i}(\tau+1)=s^{\prime}}}{N\left(t-t_{0}-1\right)}.
\label{ss}
\end{equation}
Here the function $\mathbbm{1}_{\text{predicate}}$ equals one if the predicate of the two consecutive states for player $i$ is true, and zero otherwise. Therefore, Cohen’s kappa coefficient $\kappa(s,s')$ captures the transition ``flow" from state $s$ to $s'$. If a pair of non-diagonal positive coefficients $\kappa (s,s') \approx \kappa (s',s)$, this corresponds to a detailed balance, otherwise there is a net transition flow between the two states. Note that diagonal elements of the matrix are for self-loops, where the system could settle down.

Fig.~\ref{fig:Kappa_density} shows both the matrices and state density distributions for the first discontinuous phase transition by fixing $\rho= 0.332$ [see Fig.~\ref{fig:Result_PM}(a,b)], where the subplots (a,b) and (c,d) are for the high and low cooperation cases, respectively. 
In the high cooperation shown in Fig.~\ref{fig:Kappa_density}(a,b), the state $s_5$ dominates with the self-loop $s_5\leftrightarrow s_5$. 
In this case, the value of $Q_{s_{5},C}$ can be analytically obtained that well matches the numerical solution in simulations, see Appendix \ref{Appendix:PM_Q5C}. 
In contrast, Fig.~\ref{fig:Kappa_density}(c,d) shows the low cooperation case, where a variety of self-loops coexist with flows among different states, including $s_0\leftrightarrow s_0$, $s_0\leftrightarrow s_1$, $s_2\leftrightarrow s_2$, $s_2\leftrightarrow s_3$, $s_3\leftrightarrow s_4$, among others. State transitions occur between adjacent states, which contrasts with the single-state transition pattern observed at high cooperation levels. Additionally, we notice that there is no transition between states $s_1$ and $s_2$, while transitions are possible between states $s_2$, $s_3$ and $s_4$. This behavior contributes to the irreversibility of highly defection-prone states, ultimately leading to a decrease in the system's cooperation level.
	
\section{Robustness}\label{sec:robustness}
\subsection{Adaptive mixing}\label{sec:AM}

\begin{figure}[htbp]
\centering
\includegraphics[width=1.0\linewidth,center]{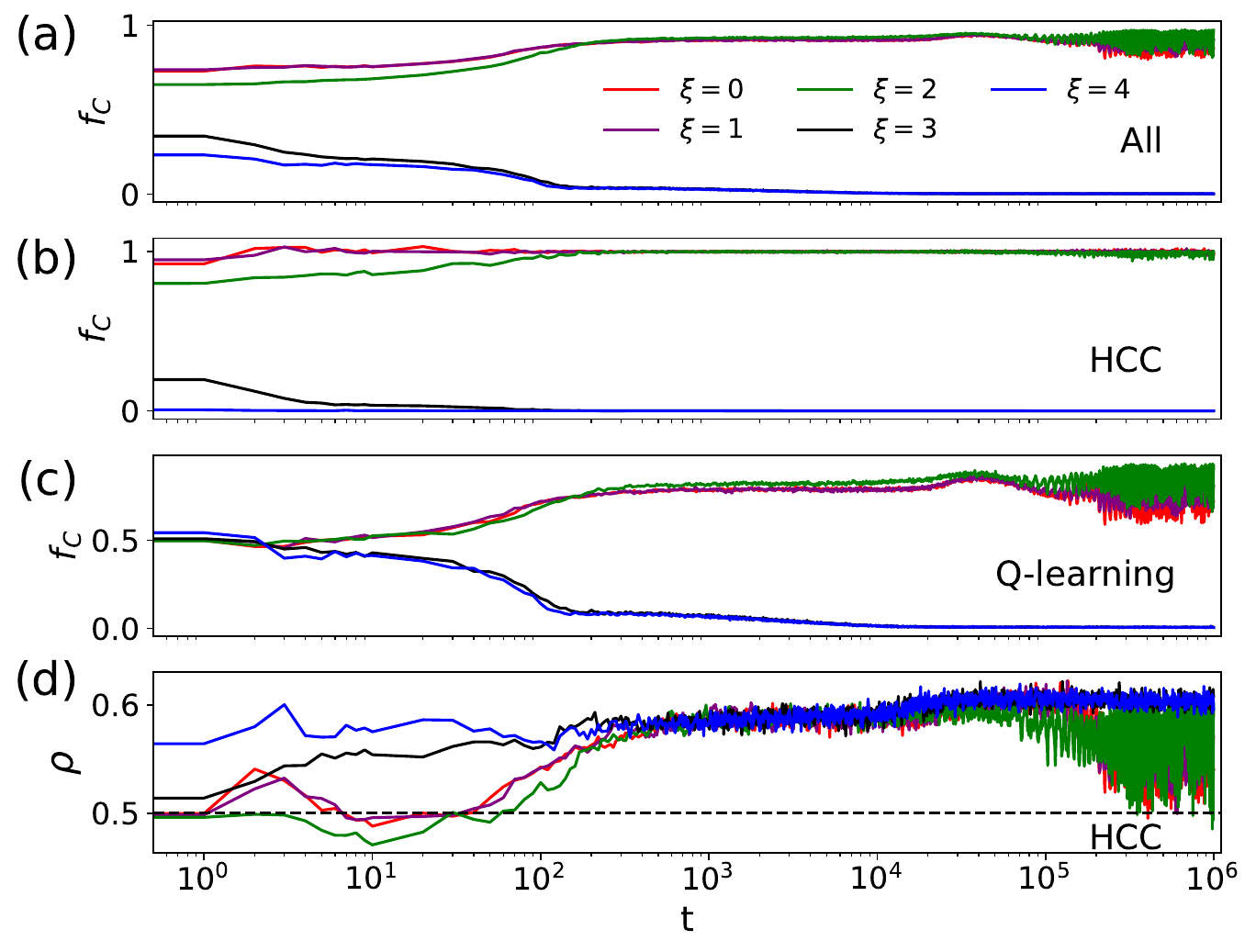}
\caption{\textbf{Evolution of cooperation with adaptive mixing}.
The cooperation level $f_C$ with different thresholds $\xi$ for the entire population (a), hard conditional cooperators (b), and players within Q-learning mode (c). 
(d) Time evolution of mode fractions for HCC.
}\label{fig:AM}
\end{figure}

Apart from SM and PM where the proportions of the two modes are fixed from the outset, here we also consider an adaptive manner for mode mixing. The adaptive mixing (AM) is as follows: when the current reward is lower than the value from the previous round, i.e., $\Pi _{i} (t)< \Pi _{i} (t-1)$, player $i$ is unhappy and thus changes its mode. Otherwise, the individual maintains the current mode. In such a way, the two conditional cooperation modes are adaptively mixed driven by higher rewards. Fig~\ref{fig:AM} illustrates the results for AM.	

As shown in Fig~\ref{fig:AM}(a), the cases with the mild threshold ($\xi$ = 0, 1, 2) achieve high levels of cooperation, while $f_C\rightarrow 0$ for higher expectation ($\xi$ = 3, 4). Notice that although high levels of cooperation are reached with lower thresholds, these cooperation levels are also oscillatory. This is because, for players with Q-learning, defecting in a highly cooperative environment yields higher rewards. But once the cooperative environment is ruined, they turn to be cooperative once again to sustain long-term rewards. Therefore, players within Q-learning mode switch between C and D repeatedly. This is confirmed in Fig~\ref{fig:AM}(c), where much stronger oscillation in $f_C$ is seen for Q-learning compared to HCC mode Fig~\ref{fig:AM}(b), and the overall prevalence is shown in Fig~\ref{fig:AM}(a).

The evolution of HCC mode fractions is stabilized at $0.50\sim0.60$ Fig~\ref{fig:AM}(d), slightly higher than the Q-learning mode. Compared to the results in Fig.~\ref{fig:Result_PM}(b), the final evolution in AM is equivalent to the observations at $\rho\approx0.6$ for PM, where the cooperation is promoted within the scenarios with $\xi=0, 1, 2$, while the cooperation disappears for $\xi=3, 4$.
	
\subsection{Conditional cooperation with Fermi-function}\label{sec:softCC}

\begin{figure}[htbp]
\centering
\includegraphics[width=1.0\linewidth, center]{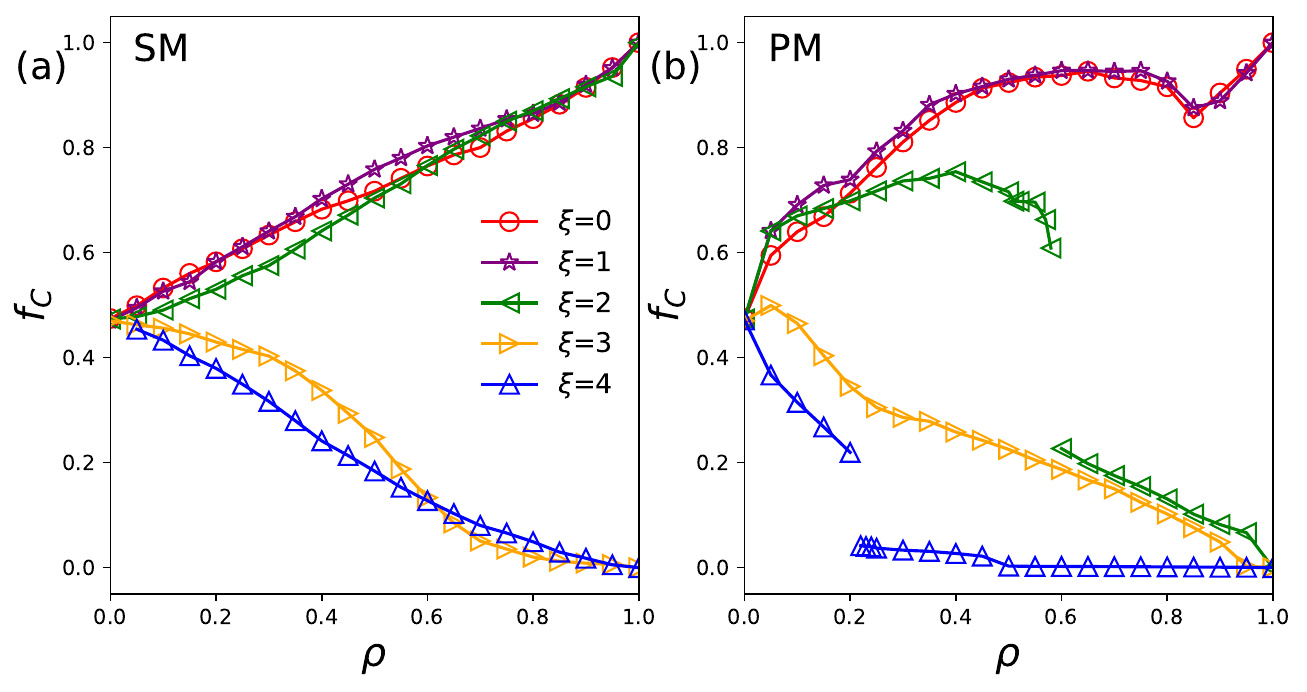}
\caption{\textbf{Results for conditional cooperation with Fermi-function}.
The overall prevalence of cooperation $f_C$, and the prevalence in two modes as a function of probability $\rho$ with both SM (a) and PM (b), for $\xi=2$ and $K=0.5$.
}\label{fig:SoftCC}
\end{figure}
	
In addition to the hard manner adopted as in Eq.(\ref{eq:hard}), we also adopt an implementation using the Fermi function, where cooperation is adopted with the following probability~\cite{SantosScale2005, Attila2015Conformity, Evolutionary1998szab}:
\begin{equation}
	\label{eq:fermi}
	p_C(i, t+1)=\frac{1}{1+\exp[(\xi-n_C(i, t))/K]},
\end{equation}
where $\xi$ is the threshold, and $K$ is a temperature-like parameter. The case of $K\rightarrow 0$ recovers to the HCC adopted above. As $K$ becomes large, people may choose cooperation even if the number of cooperators within player $i$'s neighborhood $n_C$ is smaller than the threshold $\xi$. 
Notice that, the above probabilistic version is still a mechanical implementation as HCC, players fail to adaptively revise their preference when the surroundings are changed.

Fig.~\ref{fig:SoftCC} shows the corresponding results as a function of $\rho$ for the above two mixtures for $\xi=0,...,4$ and $K=0.5$. Compared to results of HCC [i.e. Fig.~\ref{fig:Result_SM}(b) and ~\ref{fig:Result_PM}(b)], the overall trend remains robust but still differs in details. Specifically,  while the dependencies of $f_C$ on $\rho$ remain similar for the cases of $\xi=0,1,2,4$, the non-monotonic trend disappears for $\xi=3$. The reason can be attributed to the fact that even though Q-learning players try to cooperate to meet the requirements of conditional cooperators, the probabilistic manner, however, discourages the choice of Q-learning players in cooperation, as conditional cooperators with Fermi function may still defect.

Comparatively, the difference in PM is even stronger. While the curves are largely the same for the two cases of $\xi=0, 1$,  the abrupt transition is now seen for $\xi=2$ instead of $\xi=3$ as in the case of HCC. This is because the cooperation probability with the form of Eq. (\ref{eq:fermi}) with $\xi=2$ [e.g. $P_C (n_C=2)=0.5$ and $P_C(n_C=3)\approx0.88$] is comparable to the case of $\xi=3$ within HCC [$P_C (n_C=2)=0$ and $P_C(n_C=3)=1$].
Though the transition is of a single jump, not the double jumps seen in ~\ref{fig:Result_PM}(b).
Interestingly, the transition here for $\xi=3$ instead becomes continuous, while the transition for $\xi=4$ remains discontinuous.

\section{Conclusion and discussion}
In summary, this study investigates the impact of two types of conditional cooperation on the evolution of cooperation. In the traditional framework, hard conditional cooperators (HCC) cooperate only when the number of cooperators in their neighborhood exceeds a given threshold, leading to either full cooperation or full defection depending on the population's initial conditions. In contrast, Q-learning players, representing a novel approach, exhibit a full spectrum of conditional cooperative behaviors~\cite{YiReinforcement2022,Zhao2024Emergence,shengCatalytic2024}. These include cooperating in cooperative environments, defecting in defective ones, learning to cooperate in less cooperative scenarios to sustain overall cooperation for higher long-term rewards, and occasionally free-riding in highly cooperative settings—a behavior consistent with the hump-shaped conditional cooperation observed in experiments~\cite{Fischbacher2001Are}.


We explore the evolution of cooperation within populations blending these two conditional cooperation modes, focusing on structured and probabilistic mixing schemes. The results reveal nontrivial continuous and discontinuous dependencies of cooperation on the proportion of each mode. Notably, in probabilistic mixing, the system undergoes two discontinuous phase transitions, resulting in high, low, and vanishing levels of cooperation. Analysis of typical snapshots highlights a nucleation process akin to first-order phase transitions in physics. Furthermore, the evolution of Q-values provides insights into the adaptive preferences of Q-learning players and clarifies the dynamics of cooperation.

In both mixing schemes, Q-learning players act as a ``cooperation lubricant'' in scenarios where $\xi=3$, where cooperation would otherwise fade due to the inability of hard conditional cooperators alone to sustain it. Remarkably, through Q-learning, players cooperate to achieve higher long-term rewards, ultimately fostering a significant level of cooperation in environments that would typically favor defection. Notice that researchers are recently aware of the behavioral multi-modality in humans, where individuals opt for different behavioral modes at various scenarios, and start to reveal new complexities therein~\cite{han2022hybrid, Ma2023emergence, chen2023outlearning, shengCatalytic2024}. 


Reinforcement learning offers novel insights into the role of conditional cooperation, equipping players with learning capabilities that reflect deep-seated human traits and are supported by neural mechanisms~\cite{suzuki2011neural}. Building on the success of reinforcement learning in game-theoretic studies~\cite{Zhang2020Understanding, Zhao2022Reinforcement, Wang2023Synergistic, Ding2023emergence, ZhaoPunish2024, Ma2023emergence, ZHENG2024Evolution, li2025cooperation, Zheng2023decoding, zheng2024decoding, Zhang2019reinforcement, Zheng2023optimal, HeQ2022}, the next critical step is to uncover experimental evidence for the decision-making logic underlying human behavior. Such insights are essential for understanding a wide range of societal issues from a reinforcement learning perspective.
	
	
\section*{Code availability}
The code for generating key results in this study is available at \href{https://github.com/chenli-lab/RL-CC}{https://github.com/chenli-lab/RL-CC}.

\section*{Acknowledgments}
This work is supported by the National Natural Science Foundation of China (Grants Nos. 12075144, 12165014), the Fundamental Research Funds for the Central Universities
(Grant No. GK202401002), and the Key Research and Development Program of Ningxia Province in China (Grant No. 2021BEB04032).

\appendix

\section{Finite size effect}\label{AppendixA}

\begin{figure}[htbp]
\centering
\includegraphics[width=0.95\linewidth]{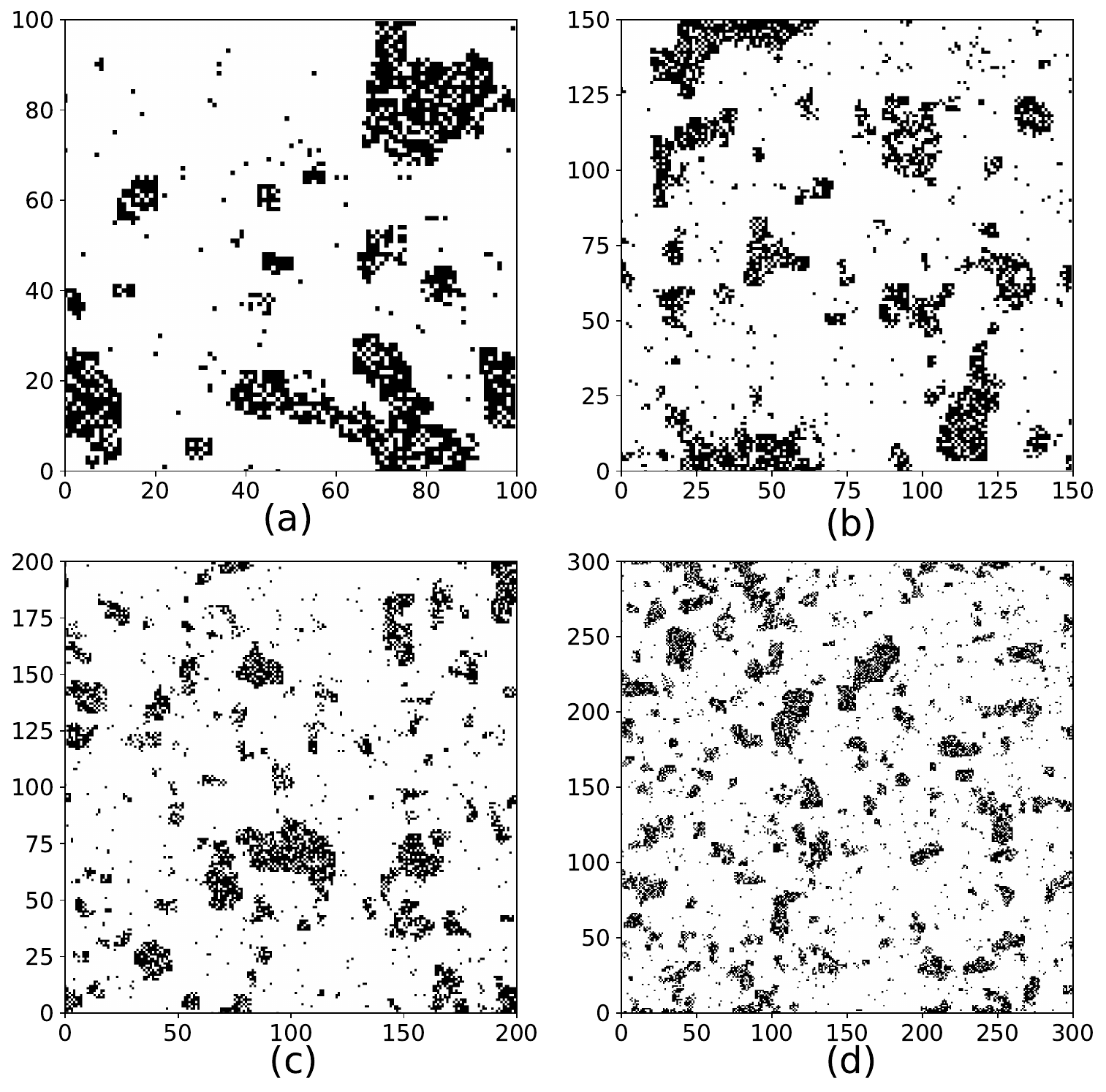}
\caption{\textbf{Typical snapshots at equilibrium for four sizes at $\rho=0.332$ with PM.}
The system size is $100\times100$, $150\times150$, $250\times250$, and $300\times300$ for (a-d), respectively. 
Other parameter: $\xi$=3.
}\label{fig:Size_snapshot}
\end{figure}
	
\begin{figure}[htbp]
\centering
\includegraphics[width=0.95\linewidth]{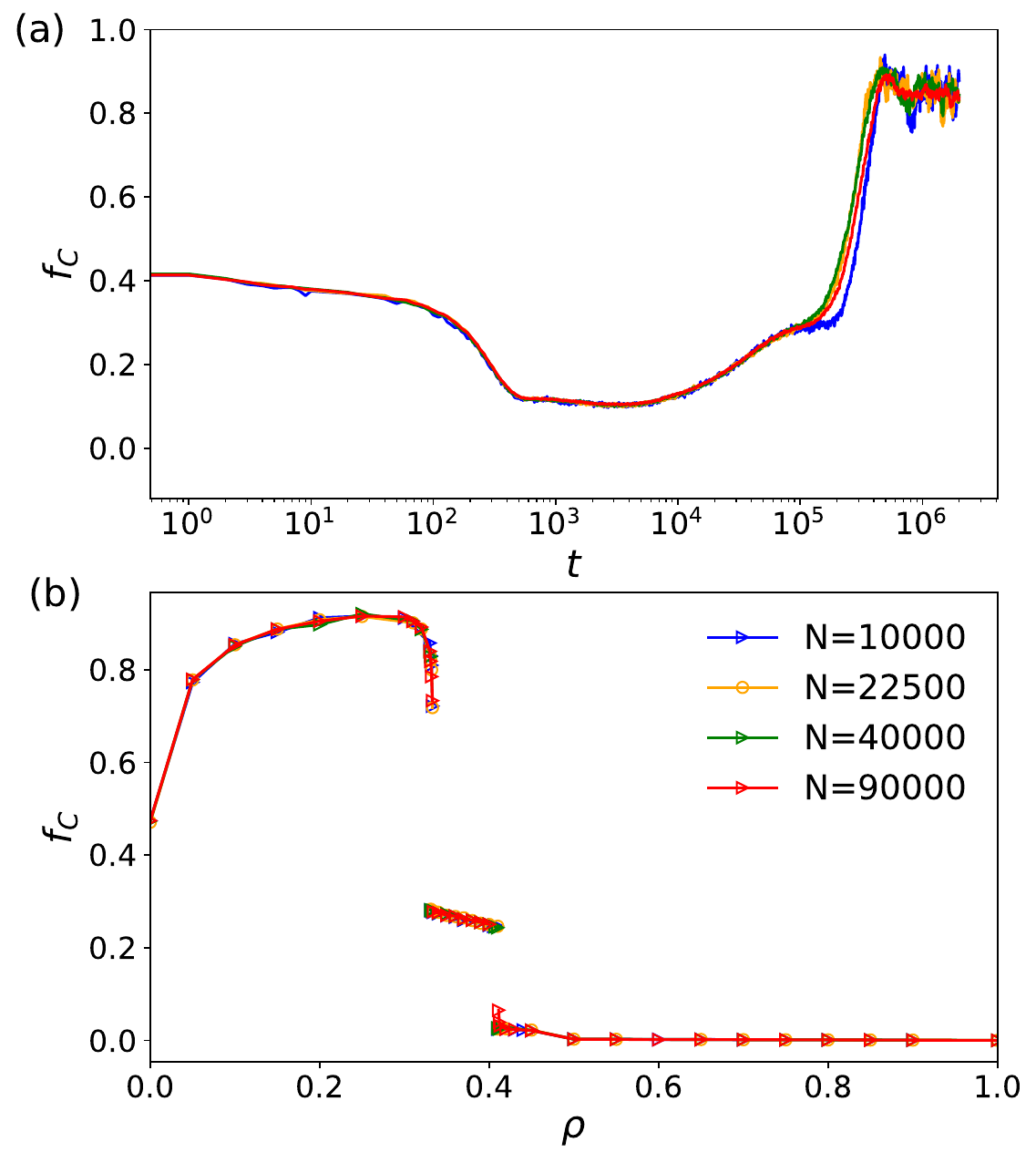}
\caption{\textbf{Impact of system size with PM.}
(a) Time series near the critical point ($\rho=0.33$) at different sizes.
(b) Cooperation phase transition $f_c$ versus $\rho$. 
Other parameter: $\xi$=3.
}\label{fig:size_effect}
\end{figure}

To explain the strong fluctuations in Fig.~\ref{fig:Result_PM}(a), we present the patterns for the four different sizes shown in Fig.~\ref{fig:Size_snapshot} with the same parameter $\rho=0.332$ around the transition point. The fluctuations are due to the formation of defector clusters, which wane and wax repeatedly. 
In the context of probabilistic mixing, if players within a defection cluster, either in CC or Q-learning mode, they remain to be defective. Conversely, players within the ``cooperation sea" also remain to be cooperators. It's the players at the interface of clusters that causes fluctuations, resulting in an oscillatory cooperation level.

Detailed observation in Fig.~\ref{fig:Size_snapshot} indicates that there is a characteristic size for these clusters. 
As the system size is increased, the fluctuations of these clusters are expected to be vanishing. This is comfirmed in the time series shown in Fig.~\ref{fig:size_effect}(a), where the oscillation in the cooperation level gradually diminishes and nearly disappears at the size of $300\times 300$. The phase transitions in Fig.~\ref{fig:size_effect}(b) show that the finite-size effect does not change the overall cooperation phase transition, where the location of the phase transition point and the cooperation levels remain unchanged for the given four sizes..

\begin{figure}[tbph]
\centering
\includegraphics[width=0.95\linewidth]{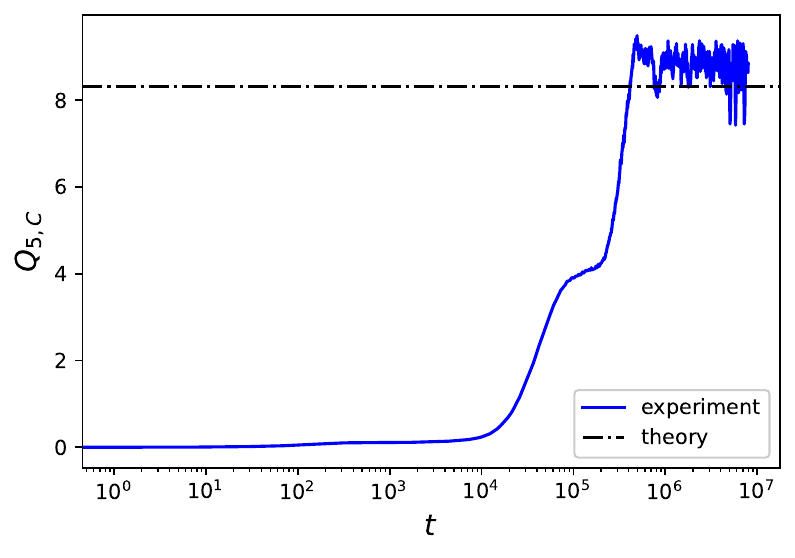}
\caption{\textbf{The evolution of $Q_{5,C}$}
The comparison between theoretical values from Eq.~(\ref{eq:theoretical_Q5C}) and the values in numerical simulations. Parameter: $\rho=0.332$.
}\label{fig:QC}
\end{figure}

\section{Estimate of $Q_{s_{5},C}$ for PM}\label{Appendix:PM_Q5C}

The expected value of $Q_{s_5,C}$ is:
\begin{equation}
    \begin{aligned}
    	   Q_{s_5,C}\to (1\!-\!\alpha)Q_{s_5,C}\!+\!(1\!-\!\frac{\epsilon }{2} )^{4}\!\times\!\alpha(\Pi _{CC}\!+\!\gamma Q_{s_5,C})\\
    	   +3\times \frac{\epsilon }{2}(1\!-\!\frac{\epsilon }{2} )^3\!\times\!\alpha(\Pi _{DC}\!+\!\gamma \max_{a'}Q_{s_4,a'}\!+\!\bigcirc (\epsilon ^2) .
    \end{aligned}
\end{equation}
 Here, the second term on the right-hand side of the above equation is the expected value of $Q_{s_5,C}$ by assuming that no exploration actions are taken. The third term illustrates how updates to the expectation occur when only one player selects the exploration action. In all other cases, where more players choose the exploration action, these effects are captured by the term $\bigcirc (\epsilon ^2)$. By further omitting $\bigcirc (\epsilon) $, we obtain the equation that satisfies the steady-state $Q_{s_5,C}$.

\begin{equation}
			Q_{s_5,C}\to (1\!-\!\alpha)Q_{s_5,C}\!+\!(1\!-\!\frac{\epsilon }{2} )^{4}\!\times\!\alpha(\Pi _{CC}+\gamma Q_{s5,C}). 
\end{equation}
We then can solve the equation and obtain:
\begin{equation}
	Q_{5,C} = \frac{(1-\frac{\epsilon }{2} )^{4}\Pi _{CC}}{1-\gamma (1-\frac{\epsilon }{2} )^4}.  
	\label{eq:theoretical_Q5C}
\end{equation}
Fig.~\ref{fig:QC} shows the theoretical solution well matches the value solution $Q_{5,C}\approx 8.3$ in the simulation.

	
\bibliography{References}
\end{document}